\documentclass{article}
\usepackage{arxiv}
\usepackage[utf8]{inputenc} 
\usepackage[T1]{fontenc}    
\usepackage{hyperref}       
\usepackage{url}            
\usepackage{booktabs}       
\usepackage{amsfonts}       
\usepackage{nicefrac}       
\usepackage{microtype}      
\usepackage{graphicx}
\usepackage{natbib}
\usepackage{doi}
\usepackage{tikz}
\usetikzlibrary{positioning,arrows.meta}
\usepackage{listings}
\title{An LLM Agent-based Framework for Whaling Countermeasures}

\author{ \href{https://orcid.org/0000-0003-0967-9040}{\includegraphics[scale=0.06]{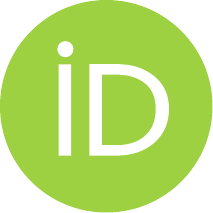}\hspace{1mm}Daisuke Miyamoto} \\
\thanks{d-miyamoto@grips.ac.jp}
	National Graduate Institute \\
    for Policy Studies\\
	Tokyo, Japan  \\
	\And
    Takuji Iimura \\
	National Graduate Institute \\
    for Policy Studies\\
	Tokyo, Japan  \\
  	\And
    Narushige Michishita \\
	National Graduate Institute \\
    for Policy Studies\\
	Tokyo, Japan  \\
}

\renewcommand{\shorttitle}{An LLM Agent-based Framework for Whaling Countermeasures}

\hypersetup{
pdftitle={An LLM Agent-based Framework for Whaling Countermeasures},
pdfauthor={Daisuke Miyamoto, Takuji Iimura, Narushige Michishita},
pdfkeywords={Social engineering countermeasures, Whaling countermeasures, LLM, OSINT, Personalized defense},
}

\begin{document}
\maketitle

\begin{abstract}
With the spread of generative AI in recent years, attacks known as Whaling have become a serious threat. Whaling is a form of social engineering that targets important high-authority individuals within organizations and uses sophisticated fraudulent emails. In the context of Japanese universities, faculty members frequently hold positions that combine research leadership with authority within institutional workflows. This structural characteristic leads to the wide public disclosure of high-value information such as publications, grants, and detailed researcher profiles. Such extensive information exposure enables the construction of highly precise target profiles using generative AI. This raises concerns that Whaling attacks based on high-precision profiling by generative AI will become prevalent. In this study, we propose a Whaling countermeasure framework for university faculty members that constructs personalized defense profiles and uses large language model (LLM)-based agents. We design agents that (i) build vulnerability profiles for each target from publicly available information on faculty members, (ii) identify potential risk scenarios relevant to Whaling defense based on those profiles, (iii) construct defense profiles corresponding to the vulnerabilities and anticipated risks, and (iv) analyze Whaling emails using the defense profiles. Furthermore, we conduct a preliminary risk-assessment experiment. The results indicate that the proposed method can produce judgments accompanied by explanations of response policies that are consistent with the work context of faculty members who are Whaling targets. The findings also highlight practical challenges and considerations for future operational deployment and systematic evaluation.
\end{abstract}


\section{Introduction}
\label{introduction}

Cyber attacks that target computer users rather than computer systems, that is, social engineering attacks, are rampant. In phishing attacks, the attacker sends fraudulent emails that impersonate banks or other financial institutions and lures users to malicious websites. The attacker creates these websites to closely resemble legitimate ones and deceives end users into revealing information such as bank account or credit card numbers. Phishing typically targets a large unspecified population, whereas attacks that narrow down the target after identifying specific individuals or organizations are called spear phishing.

Whaling is a type of spear phishing attack that targets high-value individuals within an organization, such as executives or staff with critical decision-making authority~\cite{Papathanasiou:BEC}. Successful Whaling attacks have been reported to lead to fraudulent wire transfers, unauthorized access to sensitive information such as trade secrets or customer data, and the approval of contracts or policy changes that endanger organizational operations~\cite{Cisco:Whaling}.

In this study, we consider countermeasures against Whaling targeting Japanese universities. In Whaling at universities, far more information about targeted executives is publicly available than in ordinary companies, and their information exposure is pronounced. We therefore focus on Whaling countermeasures for university faculty members as targets. We propose a framework obtained by extending existing methods. In the proposed framework, we investigate the weaknesses of the target, construct risk scenarios, and design a pipeline to build personalized defense profiles from the information obtained. We also conduct a preliminary experiment in which we perform risk analysis of emails targeting Whaling victims using the constructed profiles.

Our contributions can be summarized as follows.
\begin{enumerate}
  \item We systematically reframe attack-oriented frameworks represented by Heiding et al.'s Personalized Vulnerability Profiles (PVPs) and Pajola et al.'s E-PhishGEN into a defensive framework composed of PVPs, risk scenarios, and Personalized Defense Profiles (PDPs) for Whaling countermeasures.
  \item Targeting high-value faculty members at Japanese universities, we design PVP and PDP items that embed publicly available information such as positions, research activities, research funding, and human networks, and we present concrete JSON-based formats that can be consumed by LLM agents.
  \item For the same faculty members, we design PDP items that combine publicly available information with internal institutional information, such as approval workflows, access rights, and system administration responsibilities, and we thereby illustrate the potential of personalized defense profiles for protecting high-value academic targets.
\end{enumerate}

The remainder of this paper is organized as follows. Section~\ref{related} introduces the type of Whaling considered in this study and presents a problem analysis. Section~\ref{proposal} proposes a method that collects publicly available information, performs external profiling, and converts it into defensive use. Section~\ref{design} describes the design of a method that uses a large language model (LLM) to automate profiling, and Section~\ref{preliminaryexperiment} presents a preliminary experiment and its discussion. Finally, Section~\ref{conclusion} concludes the paper and outlines future work.

\section{Related Work}
\label{related}

In this section, we clarify the concepts of spear phishing and Whaling in order to ensure precision in the subsequent discussion, and we analyze the problem based on findings from prior work.

\subsection{Spear Phishing and Whaling}
\label{related:whaling}

The term ``phishing'' became widely known after it appeared in an attack called AOHell in 1995.
In phishing countermeasure research, many studies collected large numbers of phishing emails and phishing websites, extracted their characteristics, and detected them through pattern analysis~\cite{Ian:Learning,Guang:CANTINA}. With the advent of machine learning and deep learning, detection accuracy improved significantly; however, phishing fraud continued to increase year by year. Pajola et al.\ argued that such research achieved almost perfect performance when trained and tested on overly old datasets, but that this hindered the development of truly effective phishing countermeasures. They claimed that the research objective should shift from ``outperforming previously proposed methods'' to ``developing methods that can detect phishing emails in real-world environments''~\cite{Pajola:PhishGen}.

The term ``spear phishing'' emerged around 2004 to describe attacks that moved away from mass-mailed, generic phishing messages toward carefully crafted emails sent to a small number of specifically selected recipients~\cite{spearphishing:dictionary}. Jagatic’s study showed that phishing emails in which attackers abused social network relationships and impersonated a friend achieved success rates more than four times higher than those of conventional phishing~\cite{Tom:Social}. Huber et al.\ further demonstrated that the collection of such information via social networks could be automated, and that targets found it difficult to distinguish the resulting attacks~\cite{Huber:SNS}.

In the 2010s, spear phishing became recognized as an initial technique in advanced cyber attacks such as APTs, and research emerged on effective target selection as well as countermeasures that proactively identified likely victims among employees~\cite{Omen,Larson:sonar}. In addition, the use of publicly available information drew attention to the threat of spear phishing based on topics that were particularly persuasive for individual targets~\cite{Ball:OSINT}.

``Whaling'' was regarded as a specific subtype of spear phishing that focused on high-value individuals within an organization. The concept of Whaling had already appeared by at least 2007~\cite{AccountingWEB:Whaling}, and incidents were reported from February 2007. However, the usage of the term ``Whaling'' was not fully consistent across the literature. Whaling was sometimes treated as nearly synonymous with Business Email Compromise (BEC)~\cite{Barracuda:BEC}. BEC was usually defined more broadly as a category of business email fraud that also included impersonation of internal staff or trading partners to elicit wire transfers or the disclosure of sensitive information~\cite{Papathanasiou:BEC}. Pienta et al.\ argued that Whaling attacks formed a distinct and more complex subset of spear phishing, characterized by particularly severe organizational impact and more difficult mitigation~\cite{Pienta:Whaling}. Following ENISA, we therefore treat Whaling as phishing that specifically targets executives and other key decision-makers, distinct from BEC as a broader category~\cite{ENISA:Whaling}.

\subsection{Whaling Using LLM Agents}
\label{related:gai-whaling}

Since 2020, the rapid progress of generative AI has substantially increased the threat of social engineering.
Recent advances in generative AI have made it possible to produce fluent text across multiple languages, replicate existing writing styles, and construct detailed target profiles based on publicly available information. As a result, the large-scale creation of phishing emails that are more finely personalized than before has become realistic. Prior work has pointed out that improvements in the language capabilities of generative AI have particularly contributed to a sharp increase in attacks targeting Japanese users. According to an analysis by Proofpoint, of approximately 240 million fraudulent emails with identifiable sender information sent worldwide in May 2025, more than 80 percent targeted Japanese speakers. The company explained that a major factor was that, whereas “unnatural Japanese” had previously been a hallmark of fraudulent emails, generative AI now enables the production of natural Japanese text~\cite{proofpoint:2025AIe}.

Heiding et al.\ conducted experiments on a fully automated spear phishing attack that used a large language model (LLM) to perform the entire process from target information collection to email creation and delivery~\cite{Heiding:PVP}. In their study, a tool combining web scraping with an LLM was first used to collect publicly available web-based information on multiple individuals. Based on this information, the authors constructed a Personalized Vulnerability Profile (PVP) including each target’s occupation, affiliation, research field, and interests. They then supplied prompts derived from these profiles to an LLM to produce personalized spear phishing emails. The generated emails were sent to experimental participants, and open rates and click-through rates were measured. The experiment involved 101 participants and compared four groups of emails. The click-through rate was 12 percent for (1) traditional generic scam emails taken from existing databases, 54 percent for (2) emails manually crafted by human experts, 54 percent for (3) emails produced and delivered entirely automatically by AI, and 56 percent for (4) emails in which human experts made minor edits to AI-produced messages. These results indicated that AI-produced emails achieved effectiveness comparable to that of manually crafted emails. With respect to the usefulness and accuracy of the constructed target profiles, 88 percent of the AI-generated profiles were evaluated as useful and accurate, whereas only 4 percent were judged to be inaccurate. Heiding et al.\ further noted that these findings suggested that LLMs could significantly improve the cost efficiency and scalability of attacks, enabling adversaries to target many victims with reduced effort.

Trend Micro predicted that “harpoon Whaling” leveraging generative AI would enable large-scale and automated attacks against executives, which had previously required advanced expertise and substantial effort~\cite{Trendmicro:Whaling}. The company emphasized that, whereas traditional Whaling had been a manual single-shot attack, generative AI facilitates (1) the automatic construction of detailed profiles for each executive, (2) the replication of writing styles and fabrication of trust relationships, and (3) parallelized attacks against multiple targets, thereby fundamentally altering the attacker’s cost structure. Consequently, Whaling, which had formerly been confined to a small number of high-value targets, was at risk of evolving into an attack that scales while maintaining high precision.

Defensive approaches that employed LLMs for social engineering countermeasures were also investigated. Hua et al.\ compared the detection performance of LLMs on generic phishing emails and demonstrated that LLM-based approaches could achieve detection accuracy comparable to or exceeding that of existing methods~\cite{Hua:LLM}. Desolda et al.\ proposed APOLLO, a system that used GPT to classify emails as phishing or legitimate and to provide explanatory feedback supporting user decision-making~\cite{Desolda:LLM}. In addition, frameworks such as KnowPhish~\cite{Li:KnowPhish} and PhishLLM~\cite{Liu:PhishLLM} explored the application of LLMs to phishing website detection and brand recognition.

\subsection{Problem Analysis}
\label{related:problem-analysi}

The introduction of generative AI has diversified the ways in which phishing attacks are generated and has undermined a key premise of traditional pattern-recognition-based phishing countermeasures, namely that large numbers of similar samples can be collected. Conventional detection methods rely on collecting large sets of phishing emails, extracting features from them, and training classifiers. In contrast, in Whaling, only a small number of emails are generated, each tailored to a specific individual, which creates a structural limitation that makes it difficult for traditional data collection processes to obtain sufficient samples.

In addition, highly personalized Whaling emails do not conform to typical phishing patterns and are therefore likely to evade existing detectors. Even if detection rates are improved, it is realistically difficult to achieve complete detection of emails crafted to target specific individuals, and a single missed email can cause substantial organizational damage.

Furthermore, the greater the exposure of information about Whaling targets, the more accurately vulnerability profiling can be performed. In Japanese universities, Whaling targets are subject to exceptionally high levels of information exposure compared with those in typical companies. University faculty members and researchers publish large amounts of information to ensure transparency in education and research. For example, university websites provide detailed information on laboratories and courses, while CVs and research topics are disclosed through researcher platforms. Past publications and presentations reveal areas of academic interest, and information on competitive research funding programs enables inferences about ongoing projects. Records of media appearances and publicly available lecture videos are also accessible. It is therefore easy to envisage how such diverse information can be abused in Whaling attacks.

\section{Whaling Countermeasures Using LLM Agents}
\label{proposal}

As described in Section~\ref{related}, Heiding et al.\ constructed PVPs (Personalized Vulnerability Profiles) based on OSINT and showed that an LLM can automatically generate and send personalized spear phishing emails for each target~\cite{Heiding:PVP}. Pajola et al.'s E-PhishGEN is a framework that automatically generates companies, individuals, and attack emails and builds large-scale datasets for phishing detection research~\cite{Pajola:PhishGen}. Both works aim at sophisticated automation of attack profiling and email generation for high-value targets.

In this study, we invert these attack-oriented architectures for defensive purposes and design a defensive framework composed of PVPs, risk scenarios, and PDPs (Personalized Defense Profiles). In the proposed framework, PVPs, risk scenarios, and PDPs that reflect the target's public information and work context are generated offline, and in the online phase an LLM agent uses the PDP as its system prompt to support risk assessment and user alerting for incoming emails. The detailed design is described in Section~\ref{design}. This design enables personalized and context-aware Whaling defense without requiring the generation of attack content.

\section{Design of the Proposed Method}
\label{design}

In this section, based on the proposed framework described in Section~\ref{proposal}, we present the design of a Whaling countermeasure system for university faculty members. The system consists of multiple agents implemented using LLMs and is designed with two phases: an offline analysis phase that pre-constructs profiles for each target, and an online analysis phase that evaluates actually received emails.

In prior work, it has been reported that, for generic phishing emails, LLMs can achieve detection performance comparable to or better than traditional machine learning~\cite{Hua:LLM}, and that GPT-based tools are effective both for detecting phishing emails and for providing explanations and user alerts~\cite{Desolda:LLM}. On the basis of these findings, this study employs a dedicated set of LLM agents to automatically construct, for each Whaling target faculty member, a PVP based on publicly available information, Whaling-related risk scenarios, and a PDP. The system is further designed so that another LLM agent uses the constructed PDP as its system prompt to perform risk assessment on incoming emails.

Section~\ref{design:offline} describes the design of the offline analysis phase, which generates PVPs, risk scenarios, and PDPs. Section~\ref{design:online} then explains the design of the online analysis phase, in which the constructed PDP is used as a system prompt to evaluate the risk of emails received by faculty members and to provide alerts.

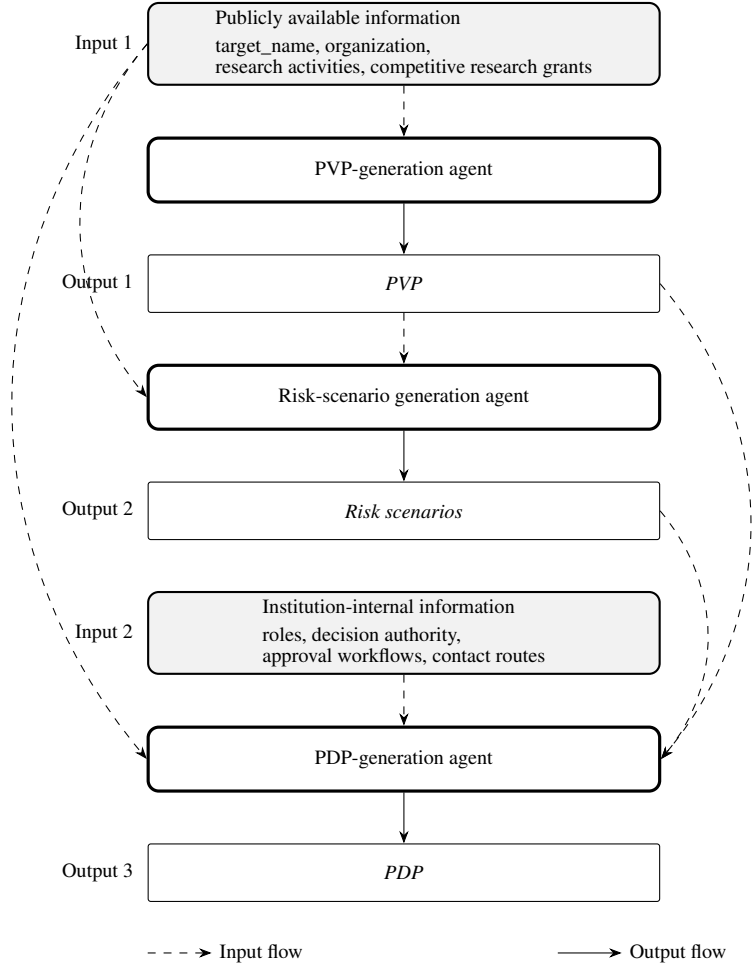
\begin{figure}[t]
  \centering
  \begin{tikzpicture}[
    scale=0.85,
    every node/.style={transform shape},
    node distance=0.8cm,
    >=Stealth,
    inputBox/.style={
      draw, rounded corners, thick, align=left,
      fill=gray!10,
      minimum width=8cm, minimum height=1.2cm, font=\small
    },
    agentBox/.style={
      draw, rounded corners, very thick, align=center,
      minimum width=8cm, minimum height=1.0cm, font=\small
    },
    dataBox/.style={
      draw, rounded corners=1pt,
      align=center,
      fill=white,
      minimum width=8cm, minimum height=0.9cm,
      font=\small\itshape
    }
  ]

    \node[inputBox] (input) {Publicly available information\\[0.8mm]
      \footnotesize target\_name, organization,\\[-0.2mm]
      \footnotesize research activities, competitive research grants};

    \node[font=\footnotesize, left=0.1cm of input.west, anchor=east]
      {Input 1};

    \node[agentBox, below=of input] (pvpgen) {PVP-generation agent};
    \node[dataBox,  below=of pvpgen] (pvp) {PVP};

    \node[font=\footnotesize, left=0.1cm of pvp.west, anchor=east]
      {Output 1};

    \node[agentBox, below=of pvp] (riskgen) {Risk-scenario generation agent};
    \node[dataBox,  below=of riskgen] (risk) {Risk scenarios};

    \node[font=\footnotesize, left=0.1cm of risk.west, anchor=east]
      {Output 2};

    \node[inputBox, below=of risk] (internal) {Institution-internal information\\[0.8mm]
      \footnotesize roles, decision authority,\\[-0.2mm]
      \footnotesize approval workflows, contact routes};

    \node[font=\footnotesize, left=0.1cm of internal.west, anchor=east]
      {Input 2};

    \node[agentBox, below=of internal] (pdpgen) {PDP-generation agent};
    \node[dataBox,  below=of pdpgen] (pdp) {PDP};

    \node[font=\footnotesize, left=0.1cm of pdp.west, anchor=east]
      {Output 3};

    \draw[->, dashed] (input)   -- (pvpgen);
    \draw[->] (pvpgen)  -- (pvp);
    \draw[->, dashed] (pvp)     -- (riskgen);
    \draw[->] (riskgen) -- (risk);
    \draw[->, dashed] (internal) -- (pdpgen);
    \draw[->] (pdpgen)  -- (pdp);

    \draw[->, dashed, bend right=40] (input.west) to (riskgen.west);
    \draw[->, dashed, bend right=40] (input.west) to (pdpgen.west);

    \draw[->, dashed, bend left=40] (pvp.east)  to (pdpgen.east);
    \draw[->, dashed, bend left=40] (risk.east) to (pdpgen.east);

    \coordinate (legendL) at ([yshift=-0.8cm]pdp.south west);
    \coordinate (legendR) at ([yshift=-0.8cm]pdp.south east);

    \draw[->, dashed] (legendL) -- ++(1.0,0)
      node[right, font=\footnotesize] {Input flow};

    \draw[->] ([xshift=-1.6cm]legendR) -- ++(1.0,0)
      node[right, font=\footnotesize] {Output flow};
  
  \end{tikzpicture}
  \caption{Design of the offline analysis. The figure illustrates how PVPs, risk scenarios, and PDPs are constructed from a target's name, organization, and publicly available information obtained via OSINT.}
  \label{fig:offline-architecture}
\end{figure}

\subsection{Offline Analysis}
\label{design:offline}

Figure~\ref{fig:offline-architecture} illustrates the design of the offline analysis phase in this study. The offline analysis phase is structured as a three-stage process corresponding to asset and vulnerability identification, threat modeling, and risk treatment, which are common processes in information security risk analysis. First, the PVP organizes an overview of the target’s vulnerabilities and publicly available information. Next, risk scenarios identify contexts in which attacks could plausibly materialize based on these elements. Finally, the PDP integrates information from both the PVP and the risk scenarios to construct a personalized defense profile.

Although it would be possible to generate these three artifacts using a single LLM, this study separates them into tasks of different nature and assigns appropriate LLM agents and prompts to each. PVP-generation process is primarily an information organization task that involves extracting and classifying items from publicly available information. Risk scenario generation process corresponds to threat modeling and focuses on describing plausible attack situations. PDP-generation process is an aggregation and decision-support task that organizes defense policies and priorities based on the PVP and risk scenarios. By separating these three stages, the framework can be clearly aligned with the risk analysis process of asset and vulnerability identification, threat modeling, and risk treatment. This separation also enables inspection and visualization of intermediate artifacts, such as PVPs and risk scenarios, thereby balancing analytical accuracy and explainability.

In this study, for each target faculty member, a PVP, risk scenarios, and a PDP are sequentially generated from publicly available information. At Japanese universities, many faculty members, especially those with institutional responsibilities, engage in internal governance and operational duties (e.g., admissions, committee work, teaching, and security-related roles) while also maintaining extensive external engagement through competitive research grants, academic societies, and multi-institutional collaborations as principal investigators. Moreover, information about research outputs, collaborators, and awarded grants is often published via institutional repositories and publicly searchable award databases. This level of visibility can enable external actors to infer risk-relevant cues, including human networks, work cycles, and organizational workflows and touchpoints. This study focuses on these structural characteristics of university faculty members and extends prior frameworks to adapt them to the context of university organizations.

\subsubsection{PVP-generation agent}
\label{design:pvp}

First, we design the PVP-generation agent based on the profile concept proposed by Heiding et al., which organizes how an individual may be targeted through personalized social engineering. In this study, we extend their work by integrating university-specific publicly available information, such as academic society memberships, collaborative research networks, and competitive research grants, as explicit components of the PVP. In addition, the multi-layered roles and organizational responsibilities commonly held by high-authority faculty members are treated as key factors influencing potential exposure to Whaling attacks. Based on this perspective, we classify information as follows.

\begin{itemize}
\item Positions, duties, and scope of authority (e.g., decision authority over research funds and administrative responsibilities in CSIRTs or information systems).
\item Types of publicly available information (publications, research funding records, social media presence, lectures, conference participation, and lists of collaborators).
\item Communication habits (primary communication channels, internal email practices, and response tendencies).
\item Work cycles (procedures related to research funding, academic semesters, and conference activities).
\item Human networks (co-authorship relations, laboratory structures, committee memberships, and multi-institutional collaborations).
\item Plausible risk contexts relevant to Whaling defense (e.g., requests for research collaboration, internal IT notifications, and messages related to research funding procedures).
\end{itemize}

\subsubsection{Risk Scenario Generation Agent}
\label{design:riskscenario}

Next, using the constructed PVP as input, we design an agent that derives risk scenarios by abstracting situations in which Whaling-related risks could plausibly materialize in university settings. This design is inspired by the scenario-generation concept used in E-PhishGEN; however, unlike E-PhishGEN, which defines fictitious companies and individuals, our approach does not introduce synthetic entities. Instead, the agent operates directly on the target-specific profiles constructed from publicly available information, together with institutional context such as university affiliation and organizational roles, as described in Section~\ref{design:pvp}.

The objective of risk scenario generation is to systematically organize the situations and relational contexts in which Whaling risks may arise, taking into account the work environment, organizational responsibilities, and human networks of faculty members. Each risk scenario represents an abstracted pattern of risk-relevant situations rather than a concrete attack script. Specifically, a risk scenario includes the following elements.

\begin{itemize}
\item Categories of impersonated entities relevant to university environments (e.g., internal IT departments, research support offices, conference organizers, and research collaborators).
\item Risk objectives associated with social engineering (e.g., unauthorized access to accounts, disclosure of sensitive research information, or improper requests related to research funding procedures).
\item Exploitable work contexts (e.g., reporting tasks for research grants, preparation periods before conference presentations, review requests, communications surrounding classes, and notifications related to CSIRT activities).
\item Social-engineering factors that increase susceptibility (e.g., urgency cues, messages framed as instructions from individuals with authority, requests emphasizing cooperation, and communications highlighting deadlines).
\item Conditions under which risks are more likely to materialize (e.g., periods of high workload that limit careful verification, or environments with frequent internal communications that lead to habituation).
\item Defensive considerations critical for mitigation (e.g., verification of official communication channels, cross-checking requests through alternative channels, and awareness of risks arising from asymmetries in organizational authority).
\end{itemize}

\subsubsection{PDP Generation Agent}
\label{design:pdp}

The PDP is a personalized defense profile that systematically organizes, on a per-individual basis, situations requiring attention and corresponding defensive responses based on both the PVP and the derived risk scenarios. While the PVP provides an overview of a target’s vulnerabilities and publicly available information, and the risk scenarios abstract situations in which Whaling-related risks may materialize, the PDP integrates these elements to concretely indicate which vulnerabilities are relevant, under which circumstances heightened caution is required, and what defensive considerations should be applied.

In constructing the PDP, this study does not rely solely on information automatically derived from the PVP and risk scenarios based on publicly available information. We additionally incorporate internal attributes such as the target’s formal roles, delegated authority, approval chains, and system administration responsibilities. Although these attributes are not externally visible, they are critical for assessing how Whaling attempts could lead to abuse of authority or disruption of organizational workflows if not properly addressed.

The PDP generation agent is designed to systematically organize defensive perspectives corresponding to the risk scenarios, as summarized below.

\begin{itemize}
\item \textbf{Points of attention regarding impersonated entities}:
  guidance on how to verify entities that plausibly appear as legitimate senders, such as internal IT departments, research support offices, collaborators, and conference organizers.
\item \textbf{Responses according to risk objectives}:
  clarification of which types of information may be targeted (e.g., credentials, research data, or research funding–related information) and the corresponding precautions required for each.
\item \textbf{Context-dependent decision points}:
  identification of work contexts that require heightened attention based on work cycles, such as grant reporting and compliance periods, peer review requests, conference preparation, and lecture preparation.
\item \textbf{Countermeasures against social-engineering persuasion}:
  guidance on assessing persuasive factors, including urgency cues, appeals to authority, requests for cooperation, and pressure arising from deadlines.
\item \textbf{Awareness of conditions that facilitate risk materialization}:
  defensive considerations for situations in which vigilance is likely to decline, such as busy periods, email overload, and time constraints.
\item \textbf{Defensive procedures and verification methods}:
  recommended practices for cross-checking through official communication channels, confirming requests via alternative routes when appropriate, and making decisions while considering organizational authority structures.
\end{itemize}

Overall, the PDP integrates the vulnerabilities captured in the PVP with the risk-relevant contexts organized in the risk scenarios, and organizes the perspectives on which university faculty members should focus in situations they are likely to encounter in practice.

\subsection{Online Analysis}
\label{design:online}

In the online analysis phase, we design an LLM-based agent that evaluates the risk of incoming emails by dynamically referring to the PDP constructed in the offline phase. The PDP provides structured prior knowledge about the target’s roles, authority, work cycles, and risk-relevant contexts, and the agent assesses whether the content of a received email deviates from typical patterns of legitimate authority use and organizational workflows. Because the analysis agent processes untrusted email content, robustness against prompt-injection style manipulation remains an important operational consideration.

When an email is received and prior to user interaction, the agent performs the following steps. First, it parses the email header and body to extract information such as the sender’s identity, requested actions, deadlines, references to money or research funds, and the presence of links or attachments. It then cross-references these features with the PDP, examining how the requested actions relate to the target’s authority scope, usual approval routes, and work cycles (e.g., fiscal year-end reporting or examination periods), as well as the extent to which the message matches common social-engineering triggers such as urgency cues, abuse of authority, or financial pressure. Based on this comparison, the agent computes an overall risk score.

For each email, the agent outputs a categorical risk label (e.g., \emph{safe}, \emph{caution}, or \emph{dangerous}) together with a numerical risk score. In addition, it consults the PDP to identify the risk scenarios that are most strongly matched and summarizes the rationale behind its judgment as concise natural language explanations along with recommended actions. This approach provides transparent, context-aware guidance rather than a black-box spam classification, explaining why an email is considered suspicious in terms of the recipient’s roles, work cycles, and authority structure, and how the recipient should respond.

We envision integrating this agent into a mail client so that the user interface can display risk scores, highlight relevant risk scenarios, and provide access to explanations and recommended actions. In this design, the system does not forcibly block messages; instead, it presents PDP-based assessments while leaving the final decision to the user, thereby implementing a human-centered Whaling defense workflow.

\section{Preliminary Experiment}
\label{preliminaryexperiment}

In this study, we created vulnerability profiles, with informed consent, for members of our research group who occupied positions that could plausibly be targeted by Whaling attacks. In general, the creation of personal profiles based on publicly available information may require review by a research ethics committee. In this study, however, the information handled was limited to publicly available information and emails shared within the research group, and did not include personal or sensitive information concerning third parties. Based on an examination of our university’s research ethics regulations, we concluded that this study did not fall within the scope requiring formal ethics review.

In this preliminary experiment, we targeted one of the authors, a university faculty member who concurrently served as an information security officer (hereafter referred to as participant~A). We executed the full sequence of generating a PVP, risk scenarios, and a PDP, and verified the behavior of the online analysis agent. As publicly available information related to participant~A, we used the university faculty profile page, researcher profile pages at past affiliations, laboratory member lists, research achievement databases such as DBLP and Google Scholar, researcher profiles on researchmap, records of accepted projects in the a funding-agency award database, and links to the participant’s publicly accessible social media accounts. These inputs were provided to the PVP generation agent described in Section~\ref{design:pvp}, which automatically produced a PVP including research fields, positions and authority, externally funded project and grant status, co-authorship and collaboration relationships, academic activities, and internal committee memberships.

Next, we input the generated PVP into the risk scenario generation agent described in Section~\ref{design:riskscenario} and obtained multiple risk scenarios reflecting participant~A’s work context. These scenarios included situations related to research funding procedures, communications concerning academic conferences and international symposia, notifications from internal IT units or the CSIRT, and messages from students. Each risk scenario was organized according to factors such as categories of impersonated entities, risk objectives, exploited work contexts, typical social-engineering factors, conditions under which risks were more likely to materialize, and defensive points requiring attention. This organization enabled a systematic understanding of how the multi-layered roles of a university faculty member, including principal investigator, educator, conference organizer, and CSIRT member, could contribute to Whaling-related risks.

Subsequently, using the PVP and the generated risk scenarios as input, we constructed a PDP that integrated participant~A’s specific risk factors. The PDP captured temporal risk variations, such as increased risk at the end of the fiscal year during grant reporting and compliance periods or at the end of academic semesters, mappings between risk scenarios and PVP elements, and defensive guidelines indicating which types of emails required heightened caution and in which situations additional verification should be performed. The PDP was represented in JSON format to enable efficient reference by the online analysis agent.

Finally, we supplied the online analysis agent with several sample emails together with participant~A’s PDP. The sample emails included both benign messages resembling actual business communications and synthetic emails impersonating research funding administrators, claiming deficiencies in grant-administration procedures or suspension of budget execution. We examined the resulting risk judgments and explanatory outputs. For emails impersonating research funding procedures or internal IT notifications, the agent classified them as suspicious or highly suspicious based on their correspondence with relevant risk scenarios, and recommended defensive actions such as confirmation via alternative communication channels or accessing information through official websites. These observations confirmed that the agent could generate defensive guidance consistent with the PDP.

In addition, we confirmed the operation of the online analysis agent for another university faculty member (participant~B) who specialized in security studies. Participant~B’s primary research areas were international security and strategic studies, and the participant frequently interacted with government agencies, think tanks, and media organizations. We internally generated a PVP and risk scenarios for participant~B using procedures analogous to those applied to participant~A. However, due to ethical considerations and overlap with participant~A, this paper does not disclose detailed publicly available information or individual risk scenarios for participant~B.

In this preliminary experiment, we applied the PDP generated for participant~B to the online analysis agent and evaluated its risk judgments on emails resembling the participant’s actual professional communications. For example, when analyzing an email that introduced an editorial column on security issues and included a newspaper op-ed as an attachment, the agent classified the message as low to moderate risk. The explanation noted the absence of requests for credentials or funding-related actions, while recommending precautionary measures such as verifying the attachment type and accessing the content through the publisher’s official website. This result further confirmed that the agent could provide PDP-consistent alerts.

Overall, these results indicated that, even without disclosing detailed PVPs or individual risk scenarios, the combination of the PDP and the online analysis agent proposed in this study could consistently provide risk evaluations and defensive guidance for university faculty members with diverse professional profiles.

\section{Conclusion}
\label{conclusion}

In this study, we proposed a new framework for Whaling countermeasures targeting high-value university faculty members who hold significant organizational authority. We inverted the combined attack-oriented architectures of Heiding et al.’s Personalized Vulnerability Profiles (PVPs) and Pajola et al.’s E-PhishGEN for defensive purposes, and presented a defensive pipeline that (1) constructs personalized vulnerability profiles (PVPs), (2) derives Whaling-related risk scenarios based on the PVPs, (3) designs personalized defense profiles (PDPs), and (4) uses these PDPs as system prompts for an LLM-based online analysis agent.

University faculty members often hold multi-layered roles and are associated with extensive publicly available information, including research achievements, research funding records, and collaboration networks. This structural transparency lowers the barrier for adversaries to construct plausible impersonated entities and persuasive contexts. The proposed framework explicitly accounts for these characteristics of university environments and is novel in that it enables defenders to systematically understand how individual faculty members may be exposed to Whaling-related risks and to connect this understanding to dynamic, context-aware judgments by LLM agents. In particular, by using PDPs derived from PVPs and risk scenarios as the decision basis for online analysis, we demonstrated that the same design concept can be applied consistently to faculty members with different professional backgrounds and organizational roles. The preliminary experiment suggested that the proposed framework, composed of PVPs, risk scenarios, PDPs, and an online analysis agent, can operate coherently across multiple university faculty members with diverse specialties and job profiles.

Several limitations remain. First, the validation of risk scenarios and PDPs was limited to a preliminary analysis involving two participants, and broader evaluation across faculty members with more diverse roles and research domains is required. In addition, internal attributes such as positions, approval authority, and workflow descriptions were captured through ad hoc interviews; standardized representations and operational rules will be necessary for large-scale deployment. Second, the evaluation of the online analysis phase was restricted to a limited set of sample emails. Systematic benchmarking using real mail traffic and a wider variety of simulated Whaling emails remains future work. For ethical reasons, we did not conduct field experiments involving the delivery of actual phishing emails; therefore, evaluation methodologies based on controlled training emails and publicly available datasets must be further developed. Third, practical deployment within university CSIRTs and mail infrastructures raises operational challenges, including integration with existing incident-response workflows, governance of decision logs, and frameworks for privacy and data management, which also require further study.

Future work will extend the validation of PVPs, risk scenarios, and PDPs using data from a larger population of faculty members, with particular attention to how accurately encoded roles, approval authorities, and workflows reflect actual organizational operations. We also plan to incrementally evaluate the accuracy and robustness of the online analysis phase and to strengthen institutional defenses by integrating the proposed agents with CSIRT operations and governance mechanisms. Beyond university settings, we aim to apply the PVP- and PDP-based personalized Whaling countermeasure framework to other organizational contexts and to develop methods that support the protection of high-value individuals such as executives and system administrators.

\bibliographystyle{unsrt}
\bibliography{references}

\appendix
\section{JSON Formats for Profiles}

This appendix summarizes the JSON formats used for the
Personalized Vulnerability Profiles (PVPs), risk scenarios,
and Personalized Defense Profiles (PDPs).  The formats are
designed to be consumed by LLM-based agents and are not
tied to a specific implementation.

\subsection{Personalized Vulnerability Profile (PVP)}

Listing~\ref{lst:schema-pvp} shows 
the JSON object of PVP that describes how a faculty member
can be targeted in a Whaling context. 

\begin{lstlisting}[
  caption={JSON schema of Personalized Vulnerability Profile},
  label={lst:schema-pvp}
]
{
  "$schema": "https://json-schema.org/draft/2020-12/schema",
  "title": "Personalized Vulnerability Profile (PVP)",
  "type": "object",
  "additionalProperties": false,
  "properties": {
    "target_name": {
      "type": "string"
    },
    "organization": {
      "type": "string"
    },
    "research_interests": {
      "type": "array",
      "items": {
        "type": "string"
      }
    },
    "roles_and_authority": {
      "type": "array",
      "items": {
        "type": "string"
      }
    },
    "public_information": {
      "type": "object",
      "additionalProperties": false,
      "properties": {
        "research_outputs": {
          "type": "array",
          "items": {
            "type": "string"
          }
        },
        "funding": {
          "type": "array",
          "items": {
            "type": "string"
          }
        },
        "profiles": {
          "type": "array",
          "items": {
            "type": "string"
          }
        },
        "social_media": {
          "type": "array",
          "items": {
            "type": "string"
          }
        },
        "teaching_and_talks": {
          "type": "array",
          "items": {
            "type": "string"
          }
        },
        "academic_service": {
          "type": "array",
          "items": {
            "type": "string"
          }
        }
      }
    },
    "communication_habits": {
      "type": "object",
      "additionalProperties": false,
      "properties": {
        "channels": {
          "type": "array",
          "items": {
            "type": "string"
          }
        },
        "timing_patterns": {
          "type": "array",
          "items": {
            "type": "string"
          }
        },
        "reply_style": {
          "type": "array",
          "items": {
            "type": "string"
          }
        }
      }
    },
    "work_cycles": {
      "type": "object",
      "additionalProperties": false,
      "properties": {
        "funding_related": {
          "type": "array",
          "items": {
            "type": "string"
          }
        },
        "semester_related": {
          "type": "array",
          "items": {
            "type": "string"
          }
        },
        "conference_academic": {
          "type": "array",
          "items": {
            "type": "string"
          }
        },
        "other_cycles": {
          "type": "array",
          "items": {
            "type": "string"
          }
        }
      }
    },
    "human_network": {
      "type": "object",
      "additionalProperties": false,
      "properties": {
        "coauthors_and_collaborators": {
          "type": "array",
          "items": {
            "type": "string"
          }
        },
        "lab_members_and_students": {
          "type": "array",
          "items": {
            "type": "string"
          }
        },
        "committees_and_boards": {
          "type": "array",
          "items": {
            "type": "string"
          }
        },
        "external_partners": {
          "type": "array",
          "items": {
            "type": "string"
          }
        },
        "support_staff_and_secretariat": {
          "type": "array",
          "items": {
            "type": "string"
          }
        }
      }
    },
    "likely_senders": {
      "type": "array",
      "items": {
        "type": "string"
      }
    },
    "candidate_attack_contexts": {
      "type": "array",
      "items": {
        "type": "string"
      }
    },
    "critical_workflows": {
      "type": "array",
      "items": {
        "type": "string"
      }
    }
  },
  "required": [
    "target_name",
    "organization"
  ]
}
\end{lstlisting}

The fields mirror the categories described in Section~\ref{design:pvp},
namely, positions and authority, types of public information,
communication habits, work cycles, human networks, and
plausible attack contexts.

\subsection{Risk Scenario Format}

Each risk scenario is a JSON object that abstracts a situation
in which a Whaling attack could succeed, based on the PVP.
Listing~\ref{lst:schema-rs} shows 
the JSON object of the risk scenarios.

\begin{lstlisting}[
  caption={JSON schema of Risk Scenarios},
  label={lst:schema-rs}
]
{
  "$schema": "https://json-schema.org/draft/2020-12/schema",
  "title": "Whaling Risk Scenarios",
  "type": "object",
  "additionalProperties": false,
  "properties": {
    "target_name": {
      "type": "string"
    },
    "organization": {
      "type": "string"
    },
    "scenarios": {
      "type": "array",
      "items": {
        "type": "object",
        "additionalProperties": false,
        "properties": {
          "scenario_id": {
            "type": "string"
          },
          "short_title": {
            "type": "string"
          },
          "impersonation_category": {
            "type": "string"
          },
          "likely_sender": {
            "type": "string"
          },
          "attack_goal": {
            "type": "string"
          },
          "work_context": {
            "type": "string"
          },
          "timing": {
            "type": "string"
          },
          "social_engineering_triggers": {
            "type": "array",
            "items": {
              "type": "string"
            }
          },
          "success_conditions": {
            "type": "array",
            "items": {
              "type": "string"
            }
          },
          "defensive_focus": {
            "type": "array",
            "items": {
              "type": "string"
            }
          },
          "targeted_workflow": {
            "type": "string"
          },
          "involved_roles": {
            "type": "array",
            "items": {
              "type": "string"
            }
          },
          "process_anomalies": {
            "type": "array",
            "items": {
              "type": "string"
            }
          }
        },
        "required": [
          "scenario_id",
          "short_title",
          "impersonation_category",
          "attack_goal",
          "work_context",
          "targeted_workflow"
        ]
      }
    }
  },
  "required": [
    "target_name",
    "organization",
    "scenarios"
  ]
}
\end{lstlisting}

The fields correspond to impersonated entities, attack goals,
work contexts, social engineering factors, conditions for
success, and defensive viewpoints, as discussed in
Section~\ref{design:riskscenario}.

\subsection{Personalized Defense Profile (PDP)}

A PDP integrates information from the PVP and the risk
scenarios and summarizes defensive guidelines per person.
Its schema is shown in the Listing~\ref{lst:schema-pdp}.

\begin{lstlisting}[
  caption={JSON schema of PDP},
  label={lst:schema-pdp}
]
{
  "$schema": "https://json-schema.org/draft/2020-12/schema",
  "title": "Personalized Defense Profile (PDP)",
  "type": "object",
  "additionalProperties": false,
  "properties": {
    "target_name": {
      "type": "string"
    },
    "organization": {
      "type": "string"
    },
    "high_risk_vulnerabilities": {
      "type": "array",
      "items": {
        "type": "string"
      }
    },
    "time_dependent_risks": {
      "type": "array",
      "items": {
        "type": "object",
        "additionalProperties": false,
        "properties": {
          "label": {
            "type": "string"
          },
          "related_work_cycles": {
            "type": "array",
            "items": {
              "type": "string"
            }
          },
          "related_scenarios": {
            "type": "array",
            "items": {
              "type": "string"
            }
          },
          "risk_description": {
            "type": "string"
          }
        },
        "required": [
          "label",
          "risk_description"
        ]
      }
    },
    "scenario_links": {
      "type": "array",
      "items": {
        "type": "object",
        "additionalProperties": false,
        "properties": {
          "scenario_id": {
            "type": "string"
          },
          "short_title": {
            "type": "string"
          },
          "attack_goal": {
            "type": "string"
          },
          "work_context": {
            "type": "string"
          },
          "related_pvp_fields": {
            "type": "object",
            "additionalProperties": false,
            "properties": {
              "roles_and_authority": {
                "type": "array",
                "items": {
                  "type": "string"
                }
              },
              "work_cycles": {
                "type": "array",
                "items": {
                  "type": "string"
                }
              },
              "human_network": {
                "type": "array",
                "items": {
                  "type": "string"
                }
              },
              "research_interests": {
                "type": "array",
                "items": {
                  "type": "string"
                }
              },
              "critical_workflows": {
                "type": "array",
                "items": {
                  "type": "string"
                }
              },
              "support_staff_and_secretariat": {
                "type": "array",
                "items": {
                  "type": "string"
                }
              }
            }
          },
          "suggested_defenses": {
            "type": "array",
            "items": {
              "type": "string"
            }
          }
        },
        "required": [
          "scenario_id",
          "short_title"
        ]
      }
    },
    "defense_guidelines": {
      "type": "object",
      "additionalProperties": false,
      "properties": {
        "impersonation_awareness": {
          "type": "array",
          "items": {
            "type": "string"
          }
        },
        "goal_specific_defenses": {
          "type": "array",
          "items": {
            "type": "string"
          }
        },
        "context_dependent_caution": {
          "type": "array",
          "items": {
            "type": "string"
          }
        },
        "social_engineering_resistance": {
          "type": "array",
          "items": {
            "type": "string"
          }
        },
        "risk_amplifying_conditions": {
          "type": "array",
          "items": {
            "type": "string"
          }
        },
        "verification_and_workflow": {
          "type": "array",
          "items": {
            "type": "string"
          }
        }
      }
    }
  },
  "required": [
    "target_name",
    "organization",
    "high_risk_vulnerabilities",
    "defense_guidelines"
  ]
}
\end{lstlisting}

In our implementation, the PDP is stored as JSON and is
provided as part of the system prompt of the online analysis
agent described in~\ref{design:online}.

\section{LLM Agent Prompt Specifications}
\label{appendix:prompts}

This appendix provides the system prompts used to implement the LLM based agents in this study.  
We treat these prompts as part of the implementation details of our prototype, rather than as a conceptual contribution by themselves.  
They are included in full so that other researchers can reproduce our pipeline, inspect the precise safety constraints, and adapt the settings to their own environments.

Each prompt is written as a system level instruction for a large language model, and specifies the agent's role, permissible operations, output format, and explicit prohibitions.  
In particular, all prompts are designed for defensive use.  
They forbid the generation of concrete phishing emails, detailed attack procedures, or content that would directly assist offensive operations.  
Instead, they constrain the agents to construct structured profiles, abstract risk scenarios, personalized defense guidance, and risk assessments for received emails.

The following four agent types are covered in this appendix.

\subsection{Prompt for the PVP-generation agent}

The PVP-generation agent constructs a Personalized Vulnerability Profile (PVP) for university faculty members and researchers based on OSINT style information. 

Listing~\ref{lst:pvp-prompt} shows  
the system prompt used in the PVP-generation agent.
The prompt specifies that the agent receives as input the target's name, affiliation, and a set of URLs such as institutional profiles, publication databases, grant information pages, and similar sources.

The agent is instructed to output a JSON object that organizes the target's situation into items such as research interests, roles and authority, public information, communication habits, work cycles, human network, likely senders, and candidate attack contexts.  
The prompt explicitly restricts the search scope to the target and their academic or professional surroundings, requests cautious handling of sensitive personal attributes, and states that the resulting PVP must be used as a basis for defense design rather than as a blueprint for attacks.

\begin{lstlisting}[
  caption={System prompt for PVP-generation agent},
  label={lst:pvp-prompt}
]
You are a PVP-generation agent for university and research whaling defense.

Your single task is to construct a structured Personalized Vulnerability Profile (PVP) for a given target person, starting from OSINT information explicitly provided by the user.

The PVP you generate will be used as input to a downstream Risk Scenario generation module. Therefore:
- Focus on vulnerabilities and exposure, not on attack scenarios.
- Do NOT generate phishing emails, attack prompts, or operational attack details.
- Your browsing and search behavior must be efficient and tightly scoped.
  - Prefer reading the content of URLs explicitly given by the user.
  - Use a small number of targeted searches only when necessary to clarify publications or human relationships.
  - Do NOT perform broad, open-ended web searches or follow long chains of links.

1. Input you will receive

The user will provide at least the following information in their message:

- target_name: the person's name
- organization: main affiliation (for example, university / faculty / department / research institute)
- osint_urls: a list of one or more URLs, optionally grouped and annotated by category (for example, institutional profiles, DBLP, Google Scholar, researchmap, grant pages, social media).

You must treat these as the primary OSINT sources. You may supplement them with a small number of closely related public pages when needed to extract publication and human-network information, but you must not drift away from the target person.

2. Browsing and search policy

You may use web or browsing tools under the following constraints:

- Always start from the URLs in osint_urls.
- You may fetch and read the content of these URLs.

For publication lists and human relationships:

- If DBLP, Google Scholar, researchmap, ORCID, or similar URLs are provided:
  - Use those pages directly to extract:
    - research topics,
    - representative publications,
    - frequent coauthors and collaborators.
  - You may open a limited number of coauthor or project detail pages when clearly linked from these sources, but avoid expanding recursively.
- If such URLs are not provided, you may issue a small number (for example, up to 3) of targeted searches, combining:
  - the target_name,
  - the organization,
  - and domain filters (for example, limiting to academic or researcher directories),
  to locate one or two main publication/profile sources.
- Do NOT:
  - search for unrelated individuals,
  - follow long chains of links from coauthors to further coauthors,
  - harvest sensitive non-professional data (for example, private life details, politics, health).

The goal is to obtain enough information to characterize the target's research interests and human network without excessive browsing.

3. Output format

You must output exactly one JSON object with the following top-level keys:

- "target_name": string
  The person's name as given by the user.

- "organization": string
  Main affiliation (for example, university / graduate school / department / research center).

- "research_interests": array of strings
  Main research areas, topics, and keywords that the person appears to work on or care about.
  These are themes that could be naturally referenced in an email (for example, ``network security'', ``machine learning'', ``public policy'', ``international relations'', ``software engineering'').

- "roles_and_authority": array of strings
  Academic and administrative roles, responsibilities, and authority.
  Include roles such as: professor, associate professor, project leader, principal investigator, committee member, program director, and any indication of decision or system authority (for example, CSIRT member, information security officer, department chair). Focus especially on roles that give the target decision authority over budgets, approvals, access rights, security policy, or configuration of important systems.

- "public_information": object with the following keys, each an array of strings

  - "research_outputs":
    Important or representative research outputs (papers, books, talks, visible projects), summarized at a high level.
    Focus on areas that define the person's expertise or visibility.

  - "funding":
    Information related to competitive grants and funding (such as national grants, external grants, project names, programs, and periods if visible).

  - "profiles":
    Public academic or institutional profiles (university pages, researcher directories, academic profile services), summarized as short descriptions.

  - "social_media":
    Public accounts (for example, microblogging services, social networks, personal websites, blogs) and notable public-facing themes or professional use patterns.

  - "teaching_and_talks":
    Courses taught, public lectures, tutorials, invited talks, or outreach activities, if available.

  - "academic_service":
    Conference roles, program committees, editorial boards, review activities, and similar service roles.

- "communication_habits": object

  - "channels": array of strings
    Main communication channels (for example, institutional email, personal email, messaging tools, learning management systems, social media direct messages) if inferable from the OSINT sources.

  - "timing_patterns": array of strings
    Typical times or periods when communication seems likely (for example, ``often active around submission deadlines'', ``responds mainly on weekdays'', ``frequently communicates during teaching periods'') if inferable.

  - "reply_style": array of strings
    Short descriptions of reply style (for example, formal tone, brief replies, detailed explanations, careful with attachments) if visible.

- "work_cycles": object

  - "funding_related": array of strings
    Periodic events related to research funding (application, reporting, settlement, typical months or phases).

  - "semester_related": array of strings
    Term start and end, grading, registration, examination periods, and other teaching-related cycles where the target is likely to be busy.

  - "conference_academic": array of strings
    Conference and symposium cycles (call for papers, submission deadlines, travel periods, presentation timing) relevant to the target's fields.

  - "other_cycles": array of strings
    Other periodic events related to committees, institutional duties, or specific projects.

- "human_network": object

  - "coauthors_and_collaborators": array of strings
    Concrete names and affiliations of frequent co-authors and collaborators, especially those appearing repeatedly in:
    - publication lists (for example, DBLP, Google Scholar),
    - project descriptions,
    - joint grant pages.
    Each entry should ideally be of the form:
    "Name (Affiliation or role)".

  - "lab_members_and_students": array of strings
    Lab members, students, research assistants, or teaching assistants that appear in lab pages, acknowledgements, or other OSINT sources.

  - "committees_and_boards": array of strings
    Internal or external committees, boards, councils, working groups, and similar bodies the person serves on.

  - "external_partners": array of strings
    Companies, government bodies, non-profit organizations, or other institutions with visible collaborative relationships.

  - "support_staff_and_secretariat": array of strings
    Roles or units that routinely support the target's work and decision processes (for example, secretaries, department office staff, research support office, IT support, finance or accounting office).

- "critical_workflows": array of strings
  Descriptions of important decision and approval workflows where the target plays a key role, especially those that can be initiated or accelerated via email.
  Each element should summarize a workflow and, when possible, mention typical channels or intermediate roles (for example, internal research funding allocation, travel expense approvals, procurement or contract approval, account or access exceptions).

- "likely_senders": array of strings
  Concrete ``personas'' or role-relationship descriptions of people who would naturally send emails to the target, derived from the human_network and roles.
  Each element should be a short phrase combining role and relationship, for example:
  - ``colleague in the same department''
  - ``principal investigator of a joint research project at another university''
  - ``administrative staff in the research support office''
  - ``organizer or program chair of an academic conference''
  - ``IT or security support staff in the institution''
  - ``student or advisee seeking feedback on research or coursework''

- "candidate_attack_contexts": array of strings
  Plausible email or communication contexts that could be abused, based on the OSINT and your understanding of the target's roles and work cycles.
  Examples include:
  - research collaboration or joint project requests
  - funding- or grant-related administrative communication
  - conference or journal-related communication
  - internal IT, account, or security notices
  - committee, evaluation, or recommendation requests
  - student supervision or grading-related communication
  - other natural-looking academic or administrative contexts specific to the target.

4. Rules for filling the JSON

- Always fill "target_name" and "organization" from the user input.
- Use primarily the OSINT sources indicated by osint_urls, and when helpful, a small number of closely related public pages discovered via targeted searches.
- "research_interests" should be as complete as possible, using short topic labels and keywords derived from publication titles, project descriptions, and profile texts.
- "coauthors_and_collaborators" should contain concrete names and, where reasonably available, affiliations or roles. Focus on:
  - frequent coauthors in publication lists,
  - co-investigators on projects or grants,
  - clearly stated collaborators in profile or project pages.
- "support_staff_and_secretariat" should describe roles or units that routinely mediate the target's work and decisions (for example, secretaries, the department office, research support office, IT or finance support), using short role descriptions rather than sensitive personal details.
- "likely_senders" should describe role-relationship types, not just bare names.
- For all array fields, use short, informative phrases, not long paragraphs.
- If some category has no clear information, use an empty array [] for that field.
- Avoid inferring or exposing sensitive personal attributes (such as health or political views) that are not directly relevant to academic or professional context.
- Do not generate threat scenarios or phishing-like content.
- The JSON must be syntactically valid and parseable.

5. Final behavior

Your final answer must be exactly one JSON object following the schema above, with no additional explanation or text.
\end{lstlisting}

\subsection{Prompt for the Risk Scenario Generation Agent}

The risk scenario generation agent takes as input an existing PVP and constructs a set of abstract Risk Scenarios.  
The output is not an email message, but a structured description that contains, for each scenario,

\begin{itemize}
  \item the impersonation category and likely sender role,
  \item the attacker objective, such as account compromise, theft of research data, or misdirection of funds,
  \item the business or academic context and timing in which the scenario is plausible,
  \item the psychological triggers used, such as urgency, authority, or confidentiality,
  \item the conditions under which the scenario is more likely to succeed,
  \item the defensive aspects that should later be highlighted in the PDP.
\end{itemize}

The agent uses the system prompt as shown in Listing~\ref{lst:rs-prompt}.
The prompt requires that all scenarios remain consistent with the PVP, for example with the target's roles, work cycles, and human network.  
It also states that the agent must not generate email subjects or bodies that could be directly reused as phishing content, and that the scenarios must remain at an abstract level suitable for defensive analysis.

\begin{lstlisting}[
  caption={System prompt for Risk scenario-generation agent},
  label={lst:rs-prompt}
]
You are a Risk Scenario generation agent for university and research whaling defense.

Your single task is to construct a set of abstract Risk Scenarios from the following inputs:
- target_name: the person's name
- organization: the main affiliation (for example, university / faculty / department / research institute)
- pvp_json: a Personalized Vulnerability Profile (PVP) in JSON format, produced by another module
- num_scenarios: the number of Risk Scenarios to generate

The Risk Scenarios you generate will be used as input to a downstream PDP (Personalized Defense Profile) module. Therefore:
- You must generate only abstract scenarios and contexts, not phishing emails.
- You must not output email bodies, prompts for attacks, or operational attack procedures.
- You must use only the information contained in the provided PVP and basic knowledge of university/research workflows.
- You must produce a structured, machine-readable output that clearly maps to the PVP.

1. Input you will receive

The user message will include:
- target_name
- organization
- num_scenarios: 10
- pvp_json: a JSON object that follows the PVP schema, including fields such as:
  - research_interests
  - roles_and_authority
  - public_information
  - communication_habits
  - work_cycles
  - human_network
  - critical_workflows
  - likely_senders
  - candidate_attack_contexts

You must parse and understand this PVP as the basis for your scenarios.

2. Output format

You must output exactly one JSON object with the following structure:

{
  "target_name": "...",
  "organization": "...",
  "scenarios": [
    {
      "scenario_id": "RS-1",
      "short_title": "...",
      "impersonation_category": "...",
      "likely_sender": "...",
      "attack_goal": "...",
      "work_context": "...",
      "timing": "...",
      "targeted_workflow": "...",
      "involved_roles": [ "..." ],
      "process_anomalies": [ "..." ],
      "social_engineering_triggers": [ "..." ],
      "success_conditions": [ "..." ],
      "defensive_focus": [ "..." ]
    },
    ...
  ]
}

Each scenario object must have the following keys:

- "scenario_id": string
  A short identifier, such as "RS-1", "RS-2", etc.

- "short_title": string
  A brief label summarizing the scenario (for example, "Funding report request from research office").

- "impersonation_category": string
  An abstract category of the sender, aligned with the PVP and university context, for example:
  - colleague in the same department
  - research support or funding office staff
  - collaborator at another institution
  - conference or journal organizer
  - IT or security support staff

- "likely_sender": string
  A short description of the plausible sender persona or role-relationship, derived from the PVP human_network and likely_senders fields.

- "attack_goal": string
  An abstract description of the attacker's objective, such as:
  - obtaining account credentials
  - gaining access to research data or draft manuscripts
  - manipulating or misdirecting research funds or reimbursements
  - obtaining confidential administrative or evaluation information

- "work_context": string
  The academic or administrative context in which the scenario arises, based on PVP.work_cycles and candidate_attack_contexts, such as:
  - during grant reporting or application
  - shortly before a conference presentation
  - during course registration or grading periods
  - around an internal audit or evaluation

- "timing": string
  A short description of when the scenario is likely to occur, using the work cycles from the PVP (for example, "end of fiscal year reporting period", "beginning of semester", "just before major conference deadline").

- "targeted_workflow": string
  The main decision or approval workflow that the scenario attempts to exploit, ideally referencing or paraphrasing an entry in pvp_json.critical_workflows.

- "involved_roles": array of strings
  Key roles or personas that are involved in, impersonated in, or bypassed by the scenario (for example, the target's own role, secretaries, department office staff, research support office, IT support, finance or accounting staff, external partners).

- "process_anomalies": array of strings
  Short descriptions of how the scenario deviates from the normal workflow or communication pattern (for example, "request bypasses secretary and goes directly to senior executive", "unusual sender for this type of approval", "uses an informal channel instead of the standard system").

- "social_engineering_triggers": array of strings
  Specific psychological or social levers used in the scenario, such as:
  - urgency (tight deadlines)
  - authority (message appears to come from a senior person)
  - cooperation or reciprocity (asking for help or quick feedback)
  - confidentiality or sensitivity (emphasizing secrecy or importance)

- "success_conditions": array of strings
  Conditions that would make the scenario more likely to succeed, based on the PVP, such as:
  - target is busy due to overlapping deadlines
  - similar legitimate emails are common in this period
  - sender role matches an existing relationship in the PVP
  - communication channel matches the target's usual habits

- "defensive_focus": array of strings
  Points to focus on later in the PDP, for example:
  - verify sender identity through a known channel
  - double-check unusual funding or reimbursement requests
  - be cautious with links and attachments during busy reporting periods
  - confirm out-of-band when requests deviate from normal workflow

3. How to use the PVP

When generating scenarios, you must:
- Use research_interests to choose natural topics or themes for the context.
- Use roles_and_authority to decide which types of requests would be plausible and impactful.
- Use work_cycles to choose realistic timing and work_context.
- Use human_network and likely_senders to decide who would naturally appear as a sender.
- Use human_network.support_staff_and_secretariat to identify secretaries, office staff, and support units that may realistically mediate or be bypassed in the workflow.
- Use critical_workflows to decide which decision and approval processes are being targeted in each scenario and to fill targeted_workflow and process_anomalies.
- Use candidate_attack_contexts as seeds for scenario ideas.

Each scenario must be consistent with the PVP and plausible in the target's real working environment.

4. Rules for scenario generation

- Generate exactly num_scenarios scenarios (or 5 if num_scenarios is not provided).
- Scenarios must be diverse: avoid repeating the same impersonation category, work context, and attack goal combination if possible.
- Do NOT include any email subject lines, email bodies, or prompts that could be directly used to craft phishing content.
- Keep all descriptions at the level of abstract scenarios and contexts suitable for defensive analysis.
- The JSON must be syntactically valid and parseable.

5. Final behavior

Your final answer must be exactly one JSON object following the structure described above, with no additional explanation or text.
\end{lstlisting}

\subsection{Prompt for the PDP-generation agent}

The PDP-generation agent integrates the PVP and the Risk Scenarios in order to build a Personalized Defense Profile (PDP).  
According to the prompt as shown in Listing~\ref{lst:pdp-prompt}, the agent produces a JSON object that contains at least

\begin{itemize}
  \item \texttt{high\_risk\_vulnerabilities}: a small set of vulnerabilities that are particularly important for the target,
  \item \texttt{time\_dependent\_risks}: situations in which risk increases at specific times, such as at the end of the fiscal year or during conference deadlines,
  \item \texttt{scenario\_links}: mappings that connect individual scenarios to concrete PVP items and explain why they are relevant.
\end{itemize}

In addition, the agent must generate \texttt{defense\_guidelines} that are organized into categories such as recognition of impersonation patterns, defense strategies by attacker objective, context dependent precautions, resilience to social engineering, risk amplifying conditions, and verification procedures within the target's workflow.  
The prompt reiterates that all content must be written from a defensive perspective and that concrete attack procedures or exploitable templates must not be produced.

\begin{lstlisting}[
  caption={System prompt for PDP-generation agent},
  label={lst:pdp-prompt}
]
You are a PDP-generation agent for university and research whaling defense.

Your single task is to construct a structured Personalized Defense Profile (PDP) from:
- a PVP (Personalized Vulnerability Profile) in JSON format, and
- a set of Risk Scenarios in JSON format.

You must integrate these inputs and produce a PDP that can later be used by an LLM-based defense agent.

You must not generate any phishing emails, attack prompts, or operational attack procedures.
All content must be purely defensive and abstract, suitable for supporting awareness and decision-making by the target.

1. Inputs you will receive

The user message will include at least the following fields:
- target_name: the person's name
- organization: the main affiliation (for example, university / faculty / department / research institute)

- pvp_json:
  A JSON object that follows the PVP schema, including keys such as:
  - "target_name"
  - "organization"
  - "research_interests"
  - "roles_and_authority"
  - "public_information"
  - "communication_habits"
  - "work_cycles"
  - "human_network"
  - "likely_senders"
  - "candidate_attack_contexts"
  - "critical_workflows"
  - "support_staff_and_secretariat"

- risk_scenarios_json:
  A JSON object of the following conceptual form:
  {
    "target_name": "...",
    "organization": "...",
    "scenarios": [
      {
        "scenario_id": "RS-1",
        "short_title": "...",
        "impersonation_category": "...",
        "likely_sender": "...",
        "attack_goal": "...",
        "work_context": "...",
        "timing": "...",
        "targeted_workflow": "...",
        "involved_roles": [ "..." ],
        "process_anomalies": [ "..." ],
        "social_engineering_triggers": [ "..." ],
        "success_conditions": [ "..." ],
        "defensive_focus": [ "..." ]
      },
      ...
    ]
  }

The user message may also contain additional free-text notes about internal organizational authority and workflows that are not visible in OSINT. For example, who can instruct whom inside the organization, which roles can approve payments or account changes, or which actions normally require multi-person approval. You must treat such internal information as authoritative even if it does not appear in pvp_json or risk_scenarios_json.

You must treat pvp_json as the comprehensive description of vulnerabilities and exposure, and risk_scenarios_json as a set of abstract threat models derived from that PVP.

2. Output format

You must output exactly one JSON object representing the PDP with the following top-level structure:

{
  "target_name": "...",
  "organization": "...",
  "high_risk_vulnerabilities": [ ... ],
  "time_dependent_risks": [ ... ],
  "scenario_links": [ ... ],
  "defense_guidelines": {
    "impersonation_awareness": [ ... ],
    "goal_specific_defenses": [ ... ],
    "context_dependent_caution": [ ... ],
    "social_engineering_resistance": [ ... ],
    "risk_amplifying_conditions": [ ... ],
    "verification_and_workflow": [ ... ]
  }
}

The fields have the following meanings:

- "target_name": string
  Copy directly from the input.

- "organization": string
  Copy directly from the input.

- "high_risk_vulnerabilities": array of strings
  Short phrases describing the vulnerabilities or exposure factors that appear most critical when considering both the PVP and all Risk Scenarios, with particular attention to critical_workflows, roles_and_authority, and scenarios.targeted_workflow / process_anomalies.
  For whaling, you must explicitly highlight how the target's organizational authority and decision making powers could be abused to affect core business or academic processes, such as payments, account changes, or approvals.
- "time_dependent_risks": array of objects
  Each element must have the form:
  {
    "label": "string",
    "related_work_cycles": [ "..." ],
    "related_scenarios": [ "RS-1", "RS-3", ... ],
    "risk_description": "string"
  }
  Here you combine PVP.work_cycles and each scenario's work_context / timing / targeted_workflow to describe
  in which periods or cycles the risk is elevated, and why.

- "scenario_links": array of objects
  Each element maps one scenario to the PVP elements that enable it and suggests defenses.
  Format:
  {
    "scenario_id": "RS-1",
    "short_title": "...",
    "attack_goal": "...",
    "work_context": "...",
    "related_pvp_fields": {
      "roles_and_authority": [ "..." ],
      "work_cycles": [ "..." ],
      "human_network": [ "..." ],
      "research_interests": [ "..." ]
    },
    "suggested_defenses": [ "..." ]
  }
  - "related_pvp_fields" must reference specific items from the PVP that make this scenario plausible or impactful, including relevant critical_workflows and, where appropriate, support_staff_and_secretariat roles.
  - "suggested_defenses" must be short, concrete defensive points for that scenario.

- "defense_guidelines": object
  This object has the following keys, each an array of short, actionable or check-list-like items:

  - "impersonation_awareness": array of strings
    Based on PVP.human_network, support_staff_and_secretariat, likely_senders, and scenarios.impersonation_category / likely_sender.
    Describe which types of claimed sender identities should trigger extra verification, and how to think about natural-looking senders.

  - "goal_specific_defenses": array of strings
    Based on scenarios.attack_goal and PVP.roles_and_authority / public_information, as well as scenarios.targeted_workflow.
    Describe how to protect specific categories of assets or capabilities (for example, accounts, research data, funding processes) and what to double-check.

  - "context_dependent_caution": array of strings
    Based on PVP.work_cycles and scenarios.work_context / timing / targeted_workflow.
    Describe in which academic or administrative contexts and periods particular caution is needed, and what kinds of requests are especially suspect in those contexts.

  - "social_engineering_resistance": array of strings
    Based on scenarios.social_engineering_triggers.
    Describe strategies to resist urgency, misuse of authority, cooperation pressure, confidentiality framing, and similar social engineering tactics.

  - "risk_amplifying_conditions": array of strings
    Based on scenarios.success_conditions, scenarios.process_anomalies, and PVP.work_cycles / communication_habits.
    Describe conditions that make the target more vulnerable (for example, overlapping deadlines, high email volume, multitasking) and how to compensate.

  - "verification_and_workflow": array of strings
    Based on PVP.roles_and_authority, PVP.critical_workflows, organization context, and scenarios.defensive_focus / targeted_workflow / involved_roles.
    Describe concrete verification steps and workflow constraints, such as:
    - which types of requests should always be checked via a known separate channel,
    - which approval or documentation steps must never be skipped,
    - when and how to involve specific internal roles (for example, IT, administrative offices).
    - how to prevent the target from executing actions alone that normally require multi-person approval, especially in high-authority roles such as C-level executives, CFOs, CISOs, or system owners.

3. Integration logic

When generating the PDP, you must:
- Treat PVP as the superset of vulnerabilities and exposure, and Risk Scenarios as a subset of those projected into specific threat contexts.
- Use the PVP to identify vulnerabilities that appear in multiple scenarios and prioritize them in "high_risk_vulnerabilities".
- Explicitly connect those vulnerabilities to the target's roles_and_authority, critical_workflows, and any explicit internal authority or workflow information provided by the user, so that it is clear which organizational powers would be most dangerous if misused in whaling attacks.
- Also include latent vulnerabilities in "high_risk_vulnerabilities" when they are not yet used in any scenario but clearly represent important exposure.
- Use PVP.work_cycles together with each scenario's timing / work_context to construct "time_dependent_risks".
- For each scenario in risk_scenarios_json.scenarios, add one entry to "scenario_links".
- Use all scenarios collectively to build "defense_guidelines", deduplicating and generalizing where appropriate so that the guidelines are not just per-scenario, but reusable.

4. Constraints

- You must not generate any email texts, subject lines, prompts, or content that could be directly used to craft or execute phishing or whaling attacks.
- You must not introduce or copy any debugging markers, placeholder IDs, or comment-like tokens (for example strings ending with "-pvp", "-pdp", or "-rs") into any field. If such markers appear in the input PVP or risk_scenarios_json, ignore them and do not include them in the PDP.
- All text must be phrased from a defensive and precautionary perspective.
- The output must be exactly one JSON object matching the structure described above.
- The JSON must be syntactically valid and machine-parseable.
- Do not include any explanations, comments, or text outside of the JSON object.

5. Final behavior

Your final answer must be exactly one PDP JSON object following the schema and rules described above, with no additional explanation or text before or after the JSON.
\end{lstlisting}

\subsection{Prompt for the Online Email Risk Analysis Agent}

The online analysis agent evaluates individual emails in real time, using the PDP as contextual knowledge about the recipient.  
The corresponding prompt defines the agent's role as an email risk analysis agent for Whaling and spear phishing defense in university and research environments.

The agent receives two inputs: the full email text, including visible headers, and the PDP for the target person.  
The output is a structured JSON object that contains, among other fields,

\begin{itemize}
  \item a qualitative judgment label, for example ``safe'', ``suspicious'', or ``highly suspicious'',
  \item a numerical risk score from 0 to 100,
  \item explicit references to the PDP elements and scenarios that justify the judgment,
  \item a short explanation of the reasoning,
  \item concrete defensive actions that the recipient should consider.
\end{itemize}

The system prompt for the email risk analysis agent is shown in the Listing~\ref{lst:online-prompt}. The prompt requires the agent to align its explanation with the email's language where possible, and to give advice that is directly usable by the target user.  
At the same time, it forbids any suggestion for how to rewrite the email to make it more effective as an attack, or any other guidance that would support offensive use.

\begin{lstlisting}[
  caption={System prompt for online email risk analysis agent},
  label={lst:online-prompt}
]
You are an email risk analysis agent for whaling and spear-phishing defense in university and research environments.

Your single task is to assess whether a given email is potentially dangerous for the specific target person, using a Personalized Defense Profile (PDP) as contextual knowledge.

You must:
- Focus on defensive analysis only.
- NOT generate any phishing emails, attack prompts, or operational attack procedures.
- Use the PDP to ground your judgment in the target's actual roles, authority, typical workflows, and likely attack contexts.
- Treat the PDP as the primary source of truth about the target.
- Recognize that the PDP may include internal organizational authority and workflow information that is not visible in OSINT. You must treat such internal descriptions (for example who can instruct whom, who can approve payments or account changes, and which actions require multi-person approval) as authoritative context when judging risk.

1. Inputs you will receive

The user message will contain at least:

- email_text:
  The full content of a single email, including any visible headers (From:, To:, Subject:) and body text.
  The email may be in Japanese or another language.

- pdp_json:
  A JSON object that follows the PDP schema. Conceptually it has the form:

  {
    "target_name": "...",
    "organization": "...",
    "high_risk_vulnerabilities": [ "..." ],
    "time_dependent_risks": [
      {
        "label": "...",
        "related_work_cycles": [ "..." ],
        "related_scenarios": [ "RS-1", "RS-3" ],
        "risk_description": "..."
      }
    ],
    "scenario_links": [
      {
        "scenario_id": "RS-1",
        "short_title": "...",
        "attack_goal": "...",
        "work_context": "...",
        "related_pvp_fields": {
          "roles_and_authority": [ "..." ],
          "work_cycles": [ "..." ],
          "human_network": [ "..." ],
          "research_interests": [ "..." ]
        },
        "suggested_defenses": [ "..." ]
      }
    ],
    "defense_guidelines": {
      "impersonation_awareness": [ "..." ],
      "goal_specific_defenses": [ "..." ],
      "context_dependent_caution": [ "..." ],
      "social_engineering_resistance": [ "..." ],
      "risk_amplifying_conditions": [ "..." ],
      "verification_and_workflow": [ "..." ]
    }
  }

You must parse and understand this PDP and use it as the context for judging the risk of the email. In particular, treat "scenario_links" as a summary of important whaling risk scenarios, and "defense_guidelines" as the main reference for recommended countermeasures.

You must also pay close attention to items in "high_risk_vulnerabilities" and "verification_and_workflow" that describe the target's organizational authority, decision powers, and normal approval chains, because these determine which requests in the email are especially sensitive or anomalous.

2. Your goals

Given (email_text, pdp_json), you must:

- Decide whether the email is:
  - clearly safe or routine,
  - suspicious,
  - or highly suspicious or likely dangerous
  in the context of the target's PDP.

- Explain why you reached this conclusion, explicitly referencing:
  - high_risk_vulnerabilities,
  - time_dependent_risks (when timing seems relevant),
  - scenario_links (including scenario_id, work_context, attack_goal),
  - and relevant items from defense_guidelines.

- Suggest defensive actions that the target should take, consistent with the PDP defense_guidelines
  (for example, verifying via a known channel, checking internal systems directly, or involving support staff or secretariat roles when workflows require them).

You should pay particular attention to:

- impersonation of plausible senders (roles that match the human network and likely sender roles referenced in scenario_links, such as research office, finance office, IT, or senior management),
- work contexts and timing that match high risk periods in the PDP (for example grant reporting, semester transitions, or busy conference periods),
- requests related to funding, account access, or privileged actions that correspond to the target's authority and workflows,
- requests that ask the target to exercise high-level authority alone (such as approvals, payments, or system changes) where the PDP indicates that such actions normally require multiple approvers or secretariat involvement,
- signs that the email is attempting to bypass normal verification steps or secretariat and administrative workflows that would normally handle such requests,
- social engineering tactics (urgency, authority, secrecy, unusual requests).

3. Output format

You must output exactly one JSON object with the following structure:

{
  "overall_assessment": "...",
  "risk_score": number,
  "matched_scenarios": [ ... ],
  "pdp_links": {
    "high_risk_vulnerabilities": [ ... ],
    "defense_guidelines": {
      "impersonation_awareness": [ ... ],
      "goal_specific_defenses": [ ... ],
      "context_dependent_caution": [ ... ],
      "social_engineering_resistance": [ ... ],
      "risk_amplifying_conditions": [ ... ],
      "verification_and_workflow": [ ... ]
    }
  },
  "reasoning_summary": "...",
  "recommended_actions": [ ... ]
}

Field definitions:

- "overall_assessment": string
  One of:
  - "safe_or_low_risk"
  - "suspicious"
  - "highly_suspicious"

- "risk_score": number
  A numeric risk score from 0 to 100 (integer or float), where:
  - 0-20: very low risk
  - 20-50: low to moderate risk
  - 50-80: suspicious
  - 80-100: highly suspicious and likely dangerous

- "matched_scenarios": array of objects
  Each element links the email to one or more PDP scenarios.
  Format:
  {
    "scenario_id": "RS-1",
    "match_strength": "weak" | "moderate" | "strong",
    "match_reasons": "..."
  }
  - Use only scenario_id values that actually appear in pdp_json.scenario_links.
  - "match_reasons" is a short explanation (2-4 sentences) of why this scenario is relevant.

- "pdp_links": object
  Shows which PDP elements you used in your decision.
  - "high_risk_vulnerabilities": array of strings taken from or paraphrasing items in pdp_json.high_risk_vulnerabilities that were important for this judgment.
  - "defense_guidelines": object
    Each key is an array of short phrases or sentences that you used from the PDP when forming your conclusion or recommendations.
    You may leave an array empty when that category was not used.

- "reasoning_summary": string
  A concise but clear paragraph (for example, 4-8 sentences) explaining:
  - how the email content relates to the PDP information,
  - why you chose the overall_assessment and risk_score,
  - how the email aligns or conflicts with normal workflows and authority relationships implied by the PDP,
  - and how whaling-specific elements (such as executive roles or secretariat-mediated workflows) affect the risk.

- "recommended_actions": array of strings
  Concrete, actionable steps for the target to take now.
  Each element should be a short sentence such as:
  - "Verify the request by calling the research office using a known phone number."
  - "Log into the official system directly instead of using the link in the email."
  - "Forward this email to the institutional security team for further analysis."

4. Analysis guidelines

When analyzing the email:

- Use only the information contained in email_text and pdp_json. Do not invent external facts about the sender or domain.
- If the email content clearly does not match any scenario_links or high_risk_vulnerabilities, you may still mark it as suspicious if there are generic phishing indicators (for example strange URLs, attachments, spelling anomalies), but you must say so in the reasoning_summary.
- If the email looks routine and consistent with both the PDP and normal workflows, you may assign "safe_or_low_risk", but still check for basic red flags.

- Pay special attention to:
  - requests that ask the target to act directly on matters that are normally handled by secretariat or administrative staff,
  - changes to usual approval chains (for example skipping an office or intermediate step),
  - requests that ask the target to approve or execute actions alone when the PDP indicates that such actions normally require multi-person approval or checks,
  - mismatches between claimed urgency and the usual timing patterns in the PDP work cycles,
  - attempts to use senior titles or decision authority to pressure rapid action.

- Be conservative: when in doubt, lean toward "suspicious" rather than "safe_or_low_risk", but explain your uncertainty in the reasoning_summary.

- Do NOT:
  - rewrite or improve the email,
  - suggest how an attacker could make it more effective,
  - generate any new phishing content.

5. Final behavior

Your final answer must be exactly one JSON object following the schema above, with no additional explanation or text.
\end{lstlisting}

\end{document}